\documentclass[twocolumn,showpacs,preprintnumbers,elsart]{revtex4}
\usepackage{eurosym}
\usepackage{amssymb}
\usepackage{amsmath}
\usepackage{graphicx}
\usepackage{dcolumn}
\usepackage[center]{subfigure}
\usepackage{color}
\usepackage{float}

\begin{document}

\title{Anisotropic semi-vortices in dipolar spinor condensates controlled by
Zeeman splitting}
\author{Bingjin Liao$^{1}$, Shoubo Li$^{1}$, Chunqing Huang$^{1}$}
\author{Zhihuan Luo$^{2}$, Wei Pang$^{3}$, Haishu Tan$^{1}$}
\author{Boris A. Malomed$^{4,5,1}$}
\author{Yongyao Li$^{1}$}
\email{yongyaoli@gmail.com}
\affiliation{$^{1}$School of Physics and Optoelectronic Engineering, Foshan University,
Foshan 528000, China \\
$^{2}$College of Electronic Engineering, South China Agricultural
University, Guangzhou 510642, China\\
$^{3}$Department of Experiment Teaching,
Guangdong University of Technology, Guangzhou 510006, China\\
$^{4}$ Department of Physical Electronics, School of Electrical Engineering,
Faculty of Engineering, Tel Aviv University, Tel Aviv 69978, Israel.\\
$^{5}$Laboratory of Nonlinear-Optical Informatics, ITMO University, St.
Petersburg 197101, Russia }

\begin{abstract}
Spatially anisotropic solitary vortices, i.e., bright anisotropic vortex soliton (AVSs), supported by anisotropic
dipole-dipole interactions, were recently predicted in spin-orbit-coupled
binary Bose-Einstein condensates (BECs), in the form of two-dimensional
semi-vortices (complexes built of zero-vorticity and vortical components).
We demonstrate that the shape of the AVSs -- horizontal or vertical, with
respect to the in-plane polarization of the atomic dipole moments in the
underlying BEC -- may be effectively controlled by strength $\Omega $ of the
Zeeman splitting (ZS). A transition from the horizontal to vertical shape
with the increase of $\Omega $ is found numerically and explained
analytically. At the transition point, the AVS assumes the shape of an
elliptical ring. Mobility of horizontal AVSs is studied too, with a
conclusion that, with the increase of $\Omega $, their negative effective
mass changes the sign into positive via a point at which the effective mass
diverges. Lastly, we report a new species of \textit{inverted} AVSs, with
the zero-vorticity and vortex component placed in lower- and higher-energy
components, as defined by the ZS. They are excited states, with respect to
the ground states provided by the usual AVSs. Quite surprisingly, inverted
AVSs are \emph{stable} in a large parameter region.
\end{abstract}

\maketitle

\section{Introduction}

Solitons in two-dimensional (2D) settings have drawn a great deal of
interest as they exhibit properties which are not available in 1D, such as
vorticity, chirality, the possibility of the collapse and suppression of the
collapse, spatial anisotropy, etc. In usual 2D systems, all localized modes
supported by the ubiquitous cubic self-focusing nonlinearity are prone to
instability driven \ by the critical collapse in the same setting \cite%
{Berge1998,Fibich1999,Fibich-book}. 2D bright solitons with embedded
vorticity are subject to an even stronger instability against perturbations
breaking their axial symmetry \cite{Desyatnikov2005}. Therefore, the
creation of \emph{stable} 2D, and even 3D, bright solitary waves and
vortices remains a challenging problem, especially in the fields of optics
and Bose-Einstein condensates, where the self-focusing cubic nonlinearity
represents, respectively, the Kerr effect \cite{Kivshar2003} and attractive
interactions between atoms, which may be induced by the Feshbach resonance
\cite{Pollack2009,Bauer2009,Yan2013}.

A general method to stabilize solitons in 2D geometry is provided by the use
of spatially periodic potentials. In optics, such potentials can be imposed
by virtual photonic lattices in photorefractive crystals \cite{Efremidis2002}%
, and permanent lattices written in bulk silica \cite{Szameit2006}, while in
BECs similar potentials can be induced by interference patterns (optical
lattices) created by coherent laser beams illuminating the condensate \cite%
{Morsch2006}. Although the periodical potentials can provide diverse
functionalities, they limits the mobility of solitons. To preserve the full
mobility, stabilization of solitons in the 2D free space is a challenging
objective. A possible way is suggested by using self-focusing nonlinearities
weaker than cubic. Indeed, quadratic (alias second-harmonic generating)
nonlinearity does not cause collapse in 2D \cite{old,Malomed1997,Liu2000}.
However, the quadratic nonlinearity does not remove the azimuthal splitting
instability of vortex solitons \cite{Firthand1997}. Another possibility is
the use of competing cubic self-focusing and quintic self-defocusing
nonlinear terms to stabilize 2D solitons \cite%
{Dimitrevski1998,Berntson,Mihalache2006}. In this setting, all the
fundamental solitons are stable, while vortex solitons are stabilized above
a specific threshold \cite{Spain,Bob}, which was not, thus far, achieved in
the experiment.

The most straightforward means to stabilize fundamental and vortical
solitons in 2D free space is offered by the use of nonlocal nonlinearity. In
particular, nonlocal nonlinear response in plasmas is induced by heating and
ionization \cite{Litvak1975}, and the propagation of light in nematic liquid
crystals features nonlocality resulting from long-range molecular
interactions \cite{Peccianti2004}. In BECs, nonlocality originates from
isotropic long-range Van der Waals interactions between Rydberg atoms \cite%
{Heidemann2008,Maucher2011}, as well as from dipole-dipole interactions
(DDIs) between atoms or molecules possessing permanent magnetic or electric
dipole moments \cite%
{DDIreview2009,Pedri2005,Nath2008,Tikhonenkov22008,Tikhonenkov2008,Raghunandan2015,Jiasheng2015}%
, or moments induced by an external polarizing field \cite{Yongyao2013}.
Permanent quadrupole moments give rise to similar long-range interactions in
a molecular BEC \cite{we}. The latter one, DDI, features strong tunability,
which can be isotropic or anisotropic in the 2D plane by setting the dipoles
perpendicular or parallel to the plane.

A completely different approach to the creation of self-trapped fundamental
and vortex modes was proposed in Refs. \cite%
{Borovkova2011,Borovkova2011-2,Driben2014,Tian2012,Ywu2013,Kartashov2014}.
It relies on the use of defocusing nonlinearity in the $D$-dimensional
space, with the local strength growing from the center to periphery, as a
function of radius $r$, at any rate faster than $r^{D}$. It has been shown
that such setting can support extremely robust families of soliton and
solitary vortices.

Recently, an unexpected result was revealed by the analysis of a
two-component (spinor) self-attractive BEC with the linear spin-orbit (SO)
coupling between the components, \textit{viz}., the prediction of completely
stable 2D solitons of two types, namely, semi-vortices and mixed modes (MMs)
\cite{SVS1,SVS3,SVS4,Guihua2017,Yongchang,QD2017,Yongping}. The
semi-vortices are built of one zero-vorticity and one vortical components,
while the MMs mix zero and non zero vorticities in both components. These
findings contradict the common belief that any system with the pure cubic
self-focusing cannot support stable solitons in the free 2D space.

In comparison to the stabilization of 2D solitons in free space, a still
more challenging objective is creation of stable \textit{anisotropic}
solitary vortices, i.e., bright anisotropic vortex solitons (AVSs).
Obviously, the use anisotropic DDI, corresponding to the in-plane
polarization of the atomic moments, is a natural way to achieve this
purpose. However, stable AVSs supported solely by anisotropic DDIs have not
been reported, thus far. On the other hand, recent consideration of the
SO-coupled BEC in the free space, with nonlinearity represented by
anisotropic DDIs, has predicted stable AVSs of the semi-vortex type \cite%
{Xunda2016}. Similar vortex modes were predicted as gap solitons in the
simplified 2D free-space model, with kinetic terms dropped in comparison
with the strong SO coupling \cite{SOCgapsoliton}. In the latter case, the
bandgap structure was induced, instead of lattice potentials, by the
interplay of the SO coupling with the Zeeman splitting (ZS).

The anisotropy of these vortices manifests itself in deformation of their
shapes. The AVS created in the full model (which includes the kinetic
energy) features a 2D density profile resembling a slim peanut elongated
parallel to the (in-plane) orientation of dipole moments in the condensates
(we define it as axis $x$ in the 2D plane, i.e., as the \textit{horizontal
direction}). In particular, the density profile of the vortex component
features two symmetric maxima separated in the horizontal direction, see
Fig.1(a2) in Ref. \cite{Xunda2016} and the top row in Fig. \ref{SVexp}
displayed below. Therefore, we refer to AVS modes of this type as
\textquotedblleft horizontal" ones. On the other hand, the profile of AVSs
found as the gap solitons is essentially \textquotedblleft fatter" in the
vertical direction, perpendicular to the in-plane orientation of the dipole
moments, see Fig. 3(b1) in Ref. \cite{SOCgapsoliton} and the second row in
Fig. \ref{SVexp} below. In particular, the vortex component of the
gap-soliton AVSs features two density maxima separated in the vertical
direction. Accordingly, we call similarly shaped AVSs \textquotedblleft
vertical" ones. Moreover, these two types of AVSs have opposite signs of
their effective dynamical mass, negative and positive, for the horizontal
and vertical modes, respectively.

One objective of the present work is to figure out the reason for the
difference between the horizontal and vertical AVSs, and find a way to
control their shapes. Further, we aim to study mobility of the AVSs, and to
elaborate a new species of \textit{inverted} AVSs, which have their
zero-vorticity and vortex constituents placed in the higher- and
lower-energy components, as defined by the ZS. Our study is based on the
model of the spinor BEC with the in-plane polarization of atomic moments,
combining the linear SO coupling of the Rashba type, nonlinear long-range
DDI, and the ZS effect.

The paper is structured as follows. The model is introduced in Section II.
Basic numerical results for the horizontal and vertical AVSs, supported by
some analytical considerations, are reported in Section III. The mobility
problem for horizontal AVSs is considered, by means of systematic
simulations, in Section IV. The novel type of inverted AVSs (of the
horizontal type), which turn out to be excited states, in comparison with
the ground states realized by the usual AVSs, and, quite surprisingly, are
\emph{stable} in a large parameter region, are addressed in Section V. The
paper is concluded by Section VI.

\section{The model}

Under the mean-field approximation, the evolution of the spinor wave
functions of the dipolar BECs, $\psi =(\psi _{+},\psi _{-})$, is governed by
the coupled Gross-Pitaevskii equations, written here in the normalized form,
which include the SO coupling and ZS, with strength $\Omega $:
\begin{eqnarray}
i\partial _{t}\psi _{+}&=&-{\frac{1}{2}}\nabla ^{2}\psi _{+}+\lambda \hat{D}%
^{[+]}\psi _{-}-\Omega \psi _{+} \notag \\
&&+\psi _{+}\int d\mathbf{r^{\prime }}R(%
\mathbf{r}-\mathbf{r^{\prime }})(|\psi _{+}(\mathbf{r^{\prime }})|^{2}+|\psi
_{-}(\mathbf{r^{\prime }})|^{2}),  \notag \\
i\partial _{t}\psi _{-}&=&-{\frac{1}{2}}\nabla ^{2}\psi _{-}-\lambda \hat{D}%
^{[-]}\psi _{+}+\Omega \psi _{-} \notag\\
&&+\psi _{-}\int d\mathbf{r^{\prime }}R(%
\mathbf{r}-\mathbf{r^{\prime }})(|\psi _{+}(\mathbf{r^{\prime }})|^{2}+|\psi
_{-}(\mathbf{r^{\prime }})|^{2}).  \notag\\ \label{fulleq}
\end{eqnarray}%
Here operators of the SO-coupling are%
\begin{equation}
\hat{D}^{[\pm ]}=\partial _{x}\mp i\partial _{y},  \label{D}
\end{equation}
and its strength is fixed to be $\lambda \equiv 1$, by means of rescaling.
The DDI's kernel is
\begin{equation}
R(\mathbf{r}-\mathbf{r^{\prime }})={\frac{1-3\cos ^{2}\theta }{(\epsilon
^{2}+|\mathbf{r}-\mathbf{r^{\prime }}|^{2})^{3/2}}},
\end{equation}
where cutoff $\epsilon $ is determined by the confinement in the transverse
(third) dimension \cite{SOCgapsoliton}. The form of this kernel implies that
the dipoles are polarized (by an external magnetic field), as said above, in
the positive $x$ direction in 2D $\left( x,y\right) $ plane, hence $\cos
^{2}\theta \equiv (x-x^{\prime })^{2}/|\mathbf{r}-\mathbf{r^{\prime }}|^{2}$.

Stationary states are looked for as the usual form, $\psi _{\pm }(\mathbf{r}%
,t)=\phi _{\pm }(\mathbf{r})e^{-i\mu t}$, where $\phi _{\pm }$ are
stationary wave functions, and $\mu $ is a real chemical potential. A
dynamical invariant of the system is the total norm, which is proportional
to the total number of atoms in the binary BECs:
\begin{equation}
N=N_{+}+N_{-}=\int d\mathbf{r}(|\phi _{+}|^{2}+|\phi _{-}|^{2}).
\label{ppeq}
\end{equation}
It is also relevant to define the relative share of the total number of
atoms which are kept in the vortex component:%
\begin{equation}
F_{-}=N_{-}/\left( N_{+}+N_{-}\right) .  \label{F}
\end{equation}

The other dynamical invariants are the linear momentum [see Eq. (\ref%
{Momentum}) below] and total energy,
\begin{equation}
E=E_{\mathrm{K}}+E_{\mathrm{DDI}}+E_{\mathrm{SOC}}+E_{\mathrm{ZS}}~,
\end{equation}
where $E_{\mathrm{K}}$, $E_{\mathrm{DDI}}$, $E_{\mathrm{SOC}}$ and $E_{%
\mathrm{ZS}}$ are the kinetic energy and the terms representing DDI, SO
coupling and ZS, respectively:
\begin{eqnarray}
E_{\mathrm{K}}&=&{\frac{1}{2}}\int d\mathbf{r}\left( |\nabla \phi
_{+}|^{2}+|\nabla \phi _{-}|^{2}\right) ,  \notag \\
E_{\mathrm{DDI}}&=&{\frac{1}{2}}\iint d\mathbf{r}d\mathbf{r^{\prime }}\left(
|\phi _{+}(\mathbf{r})|^{2}+|\phi _{-}(\mathbf{r})|^{2}\right) R(\mathbf{r}-%
\mathbf{r^{\prime }}) \notag\\
&&\times(|\phi _{+}(\mathbf{r^{\prime }})|^{2}+|\phi _{-}(%
\mathbf{r^{\prime }})|^{2}),  \notag \\
E_{\mathrm{SOC}}&=&\int d\mathbf{r}(\phi _{+}^{\ast }\hat{D}^{[+]}\phi
_{-}-\phi _{-}^{\ast }\hat{D}^{[-]}\phi _{+}),  \notag \\
E_{\mathrm{ZS}}&=&-\int d\mathbf{r}\Omega (|\phi _{+}|^{2}-|\phi _{-}|^{2}).
\label{threeenergy}
\end{eqnarray}%
Note that the anisotropic system does not conserve the total orbital angular
momentum, but its standard definition for each component,
\begin{equation}
\langle L_{\pm }\rangle =\frac{1}{N_{\pm }}\int {\phi _{\pm }^{\ast }\hat{L}%
\phi _{\pm }d\mathbf{r},}  \label{Leq}
\end{equation}
where $\hat{L}=-i(y\partial _{x}-x\partial _{y})$ is the angular-momentum\
operator, can be used to characterize the degree of vorticity of the
components.

Bright-soliton modes of the semi-vortex type in various systems may be
produced by the simple input \cite{SVS1},
\begin{equation}
\phi _{+}^{0}=A_{+}\exp (-\alpha _{+}r^{2}),\quad \phi _{-}^{0}=A_{-}r\exp
(i\theta -\alpha _{-}r^{2}),  \label{SVguess}
\end{equation}
where $A_{\pm }$ and $\alpha _{\pm }$ are positive constants, $\phi _{+}$
and $\phi _{-}$ being the zero-vorticity (fundamental) and vortex
components, respectively. Starting from this input AVS can be produced by
the imaginary-time-integration method \cite{ITP1,ITP2,SOM}. In the following
sections, we numerically identify different types of AVSs by varying two
control parameters, \textit{viz}., the total norm of the soliton, as given
by Eq. (\ref{ppeq}), and the ZS strength, i.e., $\Omega $ in Eq. (\ref%
{fulleq}).

Note that, fixing $\Omega >0$ in Eq. (\ref{threeenergy}), ansatz (\ref%
{SVguess}) implies that the fundamental and vortical components are
lower-energy and higher-energy ones, respectively. The opposite situation,
with $\Omega <0$ (the inverted AVS) is possible too and is considered below
in Section V.

\section{Numerical and some analytical results for anisotropic semi-vortices}

\begin{figure*}[t]
{\includegraphics[width=1.8\columnwidth]{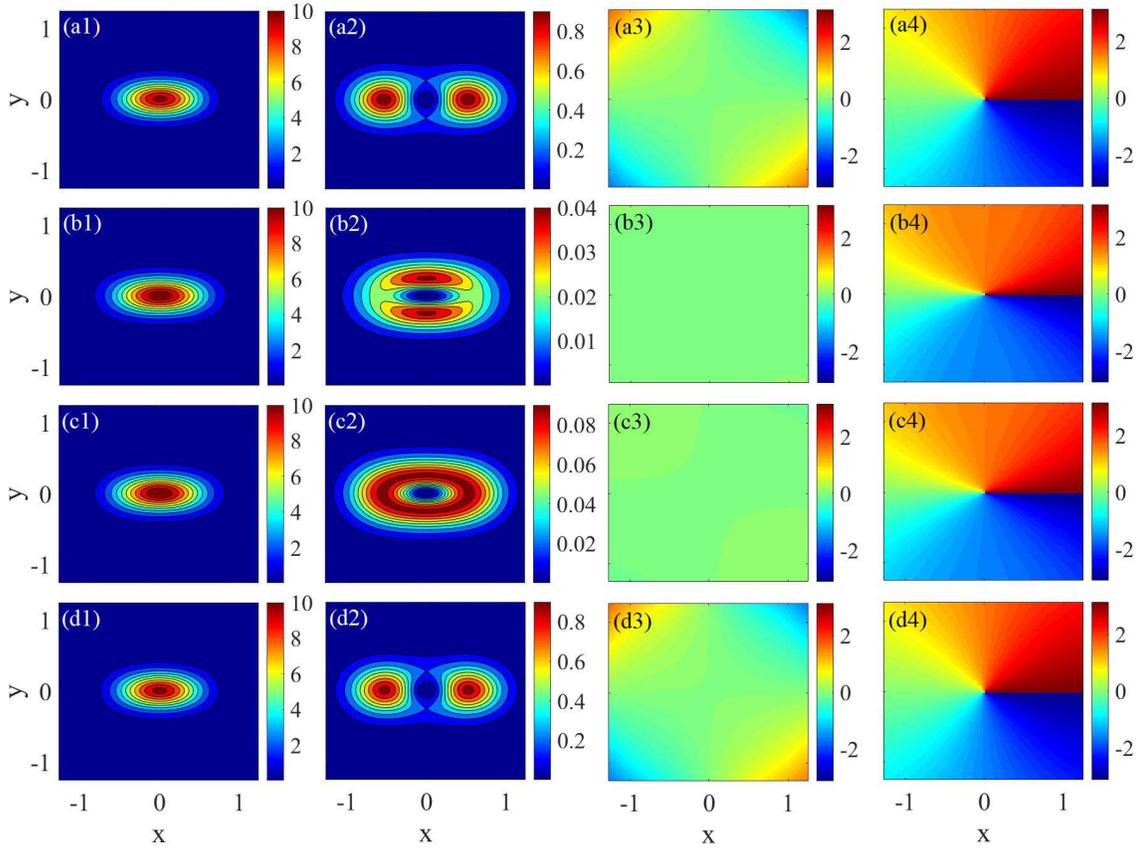}}
\caption{Displayed in the first row (from left to right) are, severally, the
density profiles of the fundamental and vortical components ($\protect\phi %
_{+}$ and $\protect\phi _{-}$, respectively), and their phase profiles, for $%
(N,\Omega )=(5,0)$, which represents a typical horizontally oriented AVS.
The second row: the same, but for $(N,\Omega )=(5,10)$, providing an example
of a vertical AVS. The third row: the same for $(N,\Omega )=(5,4.23) $,
which represents an AVS in the form of an elliptic ring. It is located
precisely at the border between horizontal and vertical modes. The fourth
row: the same for $(N,\Omega )=(5,0.0262)$, which represents a horizontal
AVS with the infinite dynamical mass (see Fig. \protect\ref{inftymass}
below).}
\label{SVexp}
\end{figure*}

\begin{figure*}[t]
{\includegraphics[width=1.2\columnwidth]{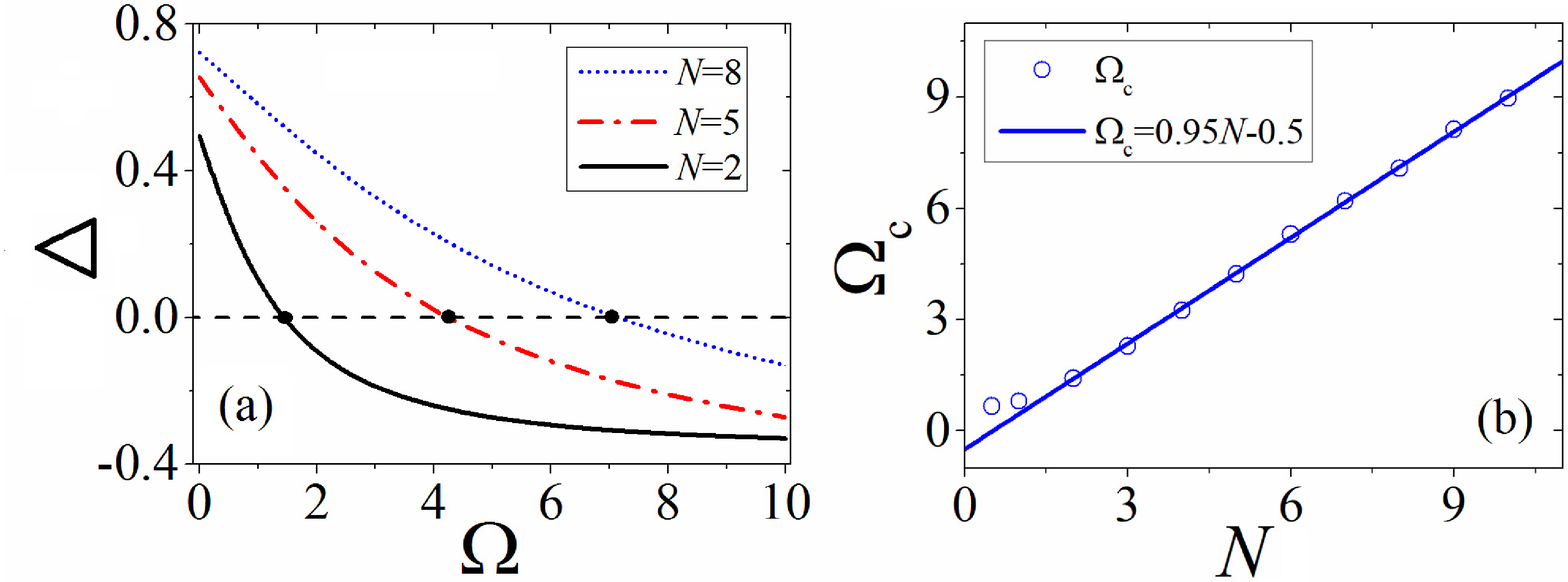}}
\caption{(a) The shape parameter, $\triangle $, defined as per Eq. (\protect
\ref{omegaC}), vs. the Zeeman-splitting strength, $\Omega$, for the total
norm $N=2$, $5$, and $8$ (black solid, red dashed-dotted, and blue dotted
curves, respectively). Black circular dots designate critical values for the
horizontal-vertical transition, $\Omega _{\mathrm{C}}(N)$. (b) $\Omega _{%
\mathrm{C}}$ vs. $N$; this dependence can be fitted by $\Omega _{\mathrm{C}%
}=0.95N-0.5$ for $N>2$. }
\label{OmegaCfig}
\end{figure*}


\begin{figure*}[t]
{\includegraphics[width=1.2\columnwidth]{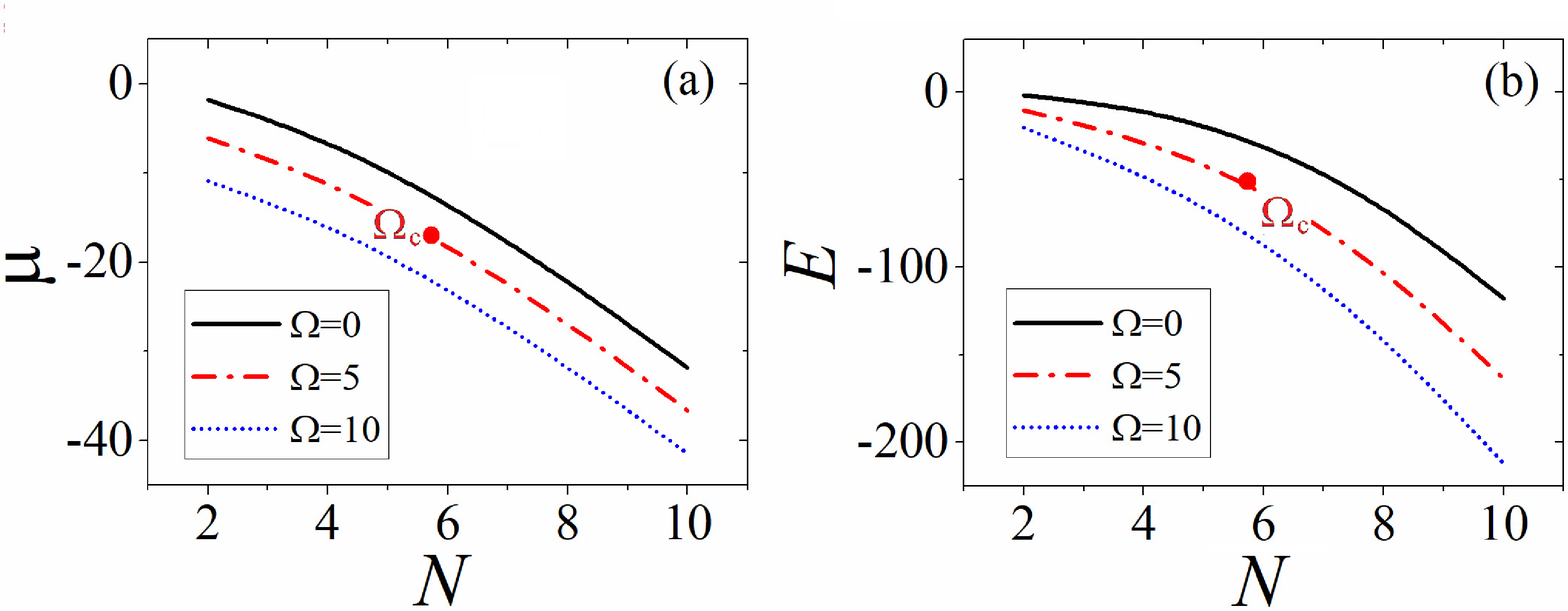}}
\caption{(a) The chemical potential, $\protect\mu $, and (b) the energy, $E$
[calculated as per Eq. (\protect\ref{threeenergy})], vs. the total norm, $N$%
, for the numerically found AVS families at different values of the
Zeeman-splitting strength, $\Omega $. The two red circle dots on the red
curves, i.e., $(N,\mu )=(5.78,-17.02)$ and $(N,E)=(5.78,-50.9)$, designate the location $\Omega _{%
\mathrm{C}}$, i.e., the boundary between the horizontal and vertical types
of the AVS.}
\label{muE}
\end{figure*}

The first and the second rows of Fig. \ref{SVexp} display typical examples
of horizontal and vertical AVSs with equal norms, generated by means of the
imaginary-time simulations at different values of the ZS strength, $\Omega $%
, starting from the same input (\ref{SVguess}). The results demonstrate that
the horizontal mode is converted into its vertical counterpart, following
the variation of $\Omega $. At $\Omega \rightarrow 0$, it is the horizontal
AVS, similar to those found in Ref. \cite{Xunda2016}, while, at $\Omega
>\Omega _{\mathrm{C}}$, it is transformed into a vertical one. To identify
the respective critical value, $\Omega _{\mathrm{C}}$, we define a shape
parameter,%
\begin{equation}
\triangle ={\frac{|\phi _{-}^{\max }(x,0)|^{2}-|\phi _{-}^{\max }(0,y)|^{2}}{%
|\phi _{-}^{\max }(x,0)|^{2}+|\phi _{-}^{\max }(0,y)|^{2}}},  \label{omegaC}
\end{equation}
where $|\phi _{-}^{\max }(x,0)|^{2}$ and $|\phi _{-}^{\max }(0,y)|^{2}$ are
local density maxima of the vortex component along the $x$ and $y$ axis,
respectively, Obviously, $\triangle >0$ and $\triangle <0$ correspond to the
horizontal and vertical shape, respectively, $\Omega _{\mathrm{C}}$ being
defined by $\triangle \equiv 0$. Figure \ref{OmegaCfig}(a) displays $%
\triangle $ as a function of $\Omega $ for different values of the total
norm, $N$, which clearly shows the transition from the horizontal to
vertical shape with the increase of $\Omega $. Figure \ref{OmegaCfig}(b)
shows the critical value, $\Omega _{\mathrm{C}}$, as a function of $N$. This
dependence may be phenomenologically fitted by
\begin{equation}
\Omega _{\mathrm{C}}=0.95N-0.5  \label{lin}
\end{equation}
for $N>2$. When $N<2$, $\Omega _{\mathrm{C}}$ deviates from this linear fit,
approaching $\Omega _{\mathrm{C}}=0.5$ at the limit of $N=0$. The shape of
the AVS precisely at $\Omega =\Omega _{\mathrm{C}}$ is displayed in the
third row of Fig. \ref{SVexp}, for $N=5$. The density pattern of the vortex
component ($\phi _{-}$) of this semi-vortex mode features an elliptical ring
with a nearly constant maximum value along it, hence it indeed may not be
identified as a horizontal or vertical mode.

The horizontal-vertical shape transition of the vortex component, and,
furthermore, the linear dependence between $\Omega _{\mathrm{C}}$ and $N$
for large values of $\Omega $ and $N$ may be qualitatively explained.
Indeed, in the limit of large $\Omega $ (strong ZS), there is a larger term,
$-\Omega $, in the condensate's chemical potential, hence solutions to Eq. (%
\ref{fulleq}) are looked for as%
\begin{equation}
\psi _{\pm }\left( x,y,t\right) =e^{i\Omega t}\Psi_{\pm} \left( x,y,t\right)
,  \label{psiPsi}
\end{equation}
where $\Psi _{\pm }$ are slowly varying functions, in comparison with $\exp
\left( i\Omega t\right) $. Then, in the lowest approximation in small
parameter $1/\Omega $, the substitution of expressions (\ref{psiPsi}) in Eq.
(\ref{fulleq}) leads to (cf. a similar approximation for different models
with the SO-coupling, developed in Ref. \cite{SVS3})%
\begin{eqnarray}
i\partial _{t}\Psi _{+}&=&-\left( {\frac{1}{2}-}\frac{\lambda ^{2}}{4\Omega
^{2}}\right) \nabla ^{2}\Psi _{+} \\
&&+\Psi _{+}\int d\mathbf{r^{\prime }}R(%
\mathbf{r}-\mathbf{r^{\prime }})(|\Psi _{+}(\mathbf{r^{\prime }})|^{2}+|\Psi
_{-}(\mathbf{r^{\prime }})|^{2}),  \label{Psi+}\notag \\
\Psi _{-}&\approx& \left( \lambda /2\Omega \right) \hat{D}^{[-]}\Psi _{+}~.
\label{Psi-}
\end{eqnarray}%
These results also imply that, at large $\Omega $, the share of the norm in
the vortex component, defined in Eq. (\ref{F}), obeys scaling $F_{-}\sim
\Omega ^{-2}$, which is confirmed by the numerically found dependences $%
F_{-}\left( \Omega \right) $ displayed below in Fig. \ref{FM}(a).

Thus, in the limit of large $\Omega $, Eq. (\ref{Psi+}) is tantamount to the
GPE for the single component ($\Psi _{+}$) in the dipolar condensate, which
gave rise to stable anisotropic zero-vorticity solitons in Ref. \cite%
{Tikhonenkov22008}, while the wave function of $\Psi _{-}$ is essentially
determined by Eq. (\ref{Psi-}). In particular, the shape of the soliton is,
naturally, stretched along the horizontal direction, to minimize the DDI
energy (making it negative for dipoles placed along a straight line,
parallel to their orientation). Further, looking at the stretched shape of
the $\psi _{+}$ component, in the second row of Fig. \ref{SVexp}, it is
obvious that it has the largest gradient, and, consequently, maxima of the
field induced according to Eqs. (\ref{Psi-}) and (\ref{D}), precisely at the
vertical positions at which the density maxima of the $\psi _{-}$ component
are observed in the figure. On the other hand, if $\Omega $ is not too
large, the $\psi _{-}$ component keeps its nonlinearity (as its norm is not
very small), and the minimization of the energy of the dipole-dipole
interaction between the moments concentrated at density maxima of $\psi _{-}$
obviously favors the horizontal shape, which is observed in the first row of
Fig. \ref{SVexp}. Finally, to estimate the critical value of $\Omega $ at
the horizontal-vertical shape-transition point, we note that the transition
is determined by the competition of the DDI and ZS energies. Looking at the
respective terms in Eq. (\ref{threeenergy}), it is easy to see that, in the
limit of large $\Omega $ and large $N$, these energy scale as $N^{2}$ and $%
\Omega N$, respectively, which explains that the transition must take place
at $\Omega _{\mathrm{C}}=\mathrm{const}\cdot N$, in agreement with Eq. (\ref%
{lin}).

Finally, overall characteristics of the families of AVSs, produced by the
numerical calculations, are provided by dependences of their chemical
potential ($\mu $) and total energy ($E$) on the total norm ($N$), which are
displayed, for different fixed values of the ZS strength ($\Omega $), in
Figs. \ref{muE}(a) and (b), respectively. In particular, it is worthy to
note that the $\mu (N)$ curves always satisfy the the Vakhitov-Kolokolov
criterion, $d\mu /dN<0$, which is a well-known necessary stability condition
for any soliton family supported by attractive interactions \cite%
{VaKo,Berge1998,Fibich-book}. Because the dominant component, $\phi _{+}$,
occupies the lower energy level split by the Zeeman effect, and its share in
the total norm grows with $\Omega $, as shown by Eq. (\ref{Psi-}), it is
easy to understand why both $\mu $ and $E$ monotonously decrease with the
increase of $\Omega $. Finally, we stress that \emph{all} the solitons are
found to be stable for $\Omega>0$.

\section{Mobility of the anisotropic semi-vortices}

\begin{figure*}[t]
{\includegraphics[width=1.6\columnwidth]{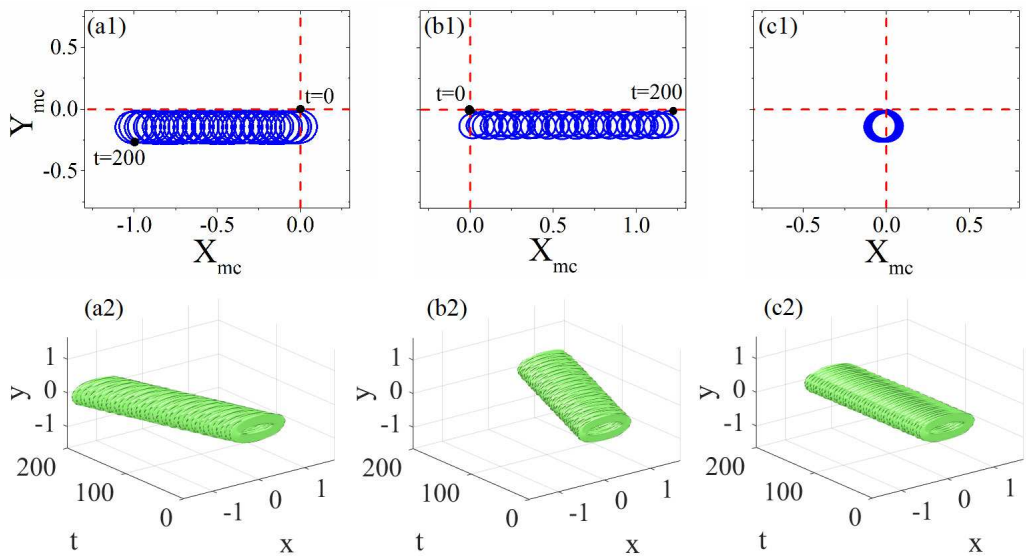}}
\caption{Typical examples of motion of horizontal AVSs with negative ($%
\Omega =0$) (a1,a2), positive ($\Omega =0.06$) (b1,b2), and infinite ($%
\Omega =0.0262$) effective mass (c1,c2) initiated by a horizontal kick, $%
\protect\eta =0.2$. The first row: trajectories of motion of the c.m. of the
fundamental component, $\protect\phi _{+}$. The second row: isosurfaces of $|%
\protect\phi _{+}|^{2}$. The total norm of the solitons in these panels is $%
N=5$.}
\label{motion}
\end{figure*}

\begin{figure*}[t]
{\includegraphics[width=1.2\columnwidth]{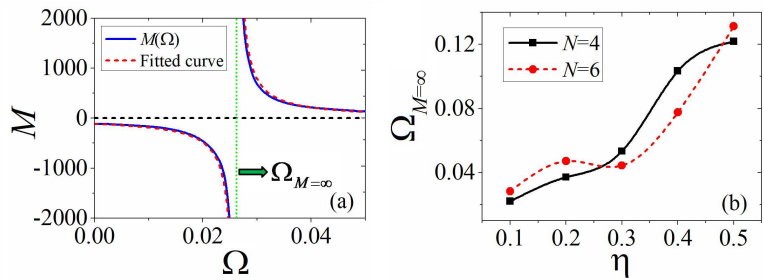}}
\caption{(a) The effective mass of the AVSs vs. $\Omega $ for norm $N=5$ and
fixed initial kick $\protect\eta =0.2$. The mass diverges at $\Omega =\Omega
_{M=\infty}=0.0262$. Dependence $M(\Omega)$ is virtually exactly fitted by $3/(\Omega-\Omega_{M=\infty})$ (the red dash curve).  (b) $\Omega _{M=\infty }$ vs. $\protect\eta $ for two values
of norm $N$ of the AVS. When $\Omega =0$, the effective mass of the soliton
is $M=-106.6$. }
\label{inftymass}
\end{figure*}

Mobility of the solitons in the system under the consideration is a
nontrivial issue, as the presence of the SO-coupling terms in Eq. (\ref%
{fulleq}) breaks its Galilean invariance \cite{SVS1}, although it conserve
the total linear momentum (unlike the nonexisting angular momentum, as
mentioned above):
\begin{equation}
\mathbf{P}=i\int d\mathbf{r}\left[ (\nabla \psi _{+}^{\ast })\psi
_{+}+(\nabla \psi _{-}^{\ast })\psi _{-}\right] .  \label{Momentum}
\end{equation}
The mobility was tested in direct simulations by applying kick $\eta $ to a
stable quiescent AVS, $\phi _{\pm }(x,y)$, in the $x$ or $y$ direction. This
correspond to simulating Eq. (\ref{fulleq}) with input
\begin{equation}
\psi _{\pm }(\mathbf{r},t=0)=\phi _{\pm }(\mathbf{r})\{e^{i\eta x},e^{i\eta
y}\},  \label{kick}
\end{equation}
the respective components of momentum (\ref{Momentum}) being $P_{x,y}=N\eta $%
. The mobility is characterized by the effective masses for the motion of
the soliton in the $x$ and $y$, directions, defined as $%
M_{x,y}=P_{x,y}/V_{x,y}$, where $V_{x,y}$ are velocities of the AVSs in the $%
x$ and $y$ directions, respectively, produced in the simulations by the
initial kick.

To define the trajectory of the kicked soliton, we introduce time-dependent
coordinates of the c.m. (center-of-mass) of its fundamental component:
\begin{equation}
\{X_{\mathrm{mc}}(t),Y_{\mathrm{mc}}(t)\}={\frac{\int \int \{x,y\}|\psi
_{+}(x,y,t)|^{2}dxdy}{\int \int |\psi _{+}(x,y,t)|^{2}dxdy}}.
\end{equation}
Indeed, the location of the c.m. of the entire two-component soliton is
dominated by the fundamental component, as the vortical one carries a small
share of the total norm, see Refs. \cite{Xunda2016,SOCgapsoliton} and Fig. %
\ref{FM}(a). The actual mobility of the kicked solitons was studied through
the shape of the c.m. trajectories, at different values of $\Omega $, as
produced by real-time simulations of Eq. (\ref{fulleq}).

In this work, we consider only the motion induced by the kick applied in the
horizontal ($x$) direction to the horizontal AVSs. Figures \ref{motion}(a,b)
illustrate the mobility of AVSs with $\Omega =0$ and $0.06$, initiated by
the kick value $\eta =+0.2$. It is seen that the resulting spiral motion may
be considered as a permanent drift of a circular trajectory of a small
radius in either negative or positive direction of $x$. The circular
component of the motion may be explained by the action of an effective
Lorentz-like force, which is a manifestation of \textit{macroscopic} SO
coupling of the AVS's intrinsic vorticity to its progressive motion \cite%
{Beijing}, although accurate consideration of this feature makes an
additional detailed analysis necessary. Lastly, it is relevant to stress the
two components of the AVS stay rigidly bound in the course of the motion.

The effective mass corresponding to the application of the kick along the $x$
direction, $M(\Omega )$, is plotted as a function \ of $\Omega $ in Fig. \ref%
{inftymass}(a). The figure reveals a surprising feature: with the increase
of $\Omega $, the effective mass changes its sign from negative to positive
via \emph{divergence} at a particular value of the ZS strength, $\Omega
=\Omega _{M=\infty }$. The red fitting curve in Fig. \ref{inftymass}(a) demonstrates that the
effective mass diverges $\sim(\Omega-\Omega_{M=\infty})^{-1}$ at $\Omega$
close to $\Omega_{M=\infty}$. Figure \ref{motion}(c) showing a typical example of real-time
evolution of AVSs with $\Omega =\Omega _{M=\infty }$ kicked by $\eta =0.2$. In
this case, the direct simulations demonstrate, in Figs. \ref{motion}(c1,c2),
that the kicked vortex performs circular motion without any progressive
drift, which may be construed as the effective immobility of the AVS.

The value of $\Omega _{M=\infty }$ depends on the total norm, as well as of the
size $\eta $ of the initial kick. This dependence is illustrated in Fig. \ref%
{inftymass}(b) by means of curves of $\Omega _{M=\infty }(\eta )$ for two
fixed values of $N$, which show that $\Omega _{M=\infty }$ increases with
the growth of $\eta $. It is worthy to note that the essential dependence of
the effective mass on the ZS strength occurs at quite small values of $%
\Omega $, therefore all the relevant AVSs are of the horizontal type [the
transition to the vertical shape happens at much higher values of $\Omega $,
see Fig. \ref{OmegaCfig}(b)]. It follows from here that all the AVSs of the
vertical type have the positive effective mass. This conclusion explains why
the horizontal AVSs in Ref. \cite{Xunda2016}, and vertical ones in Ref. \cite%
{SOCgapsoliton} feature opposite signs of their dynamical mass, as mentioned
above. Finally, the numerical results demonstrate that the effective mass
approaches the norm of the soliton, i.e., $M\rightarrow N$ in the limit of $%
\Omega\rightarrow\infty$. This conclusion is easily explained by the fact
that, in the limit of $\Omega\rightarrow\infty$, the AVS turns into a usual
single-component soliton [see Eq. (\ref{Psi-})], whose dynamical mass is
always identical to $N$.

\section{Inverted anisotropic semi-vortices, with $\Omega <0$}

\begin{figure*}[t]
{\includegraphics[width=1.8\columnwidth]{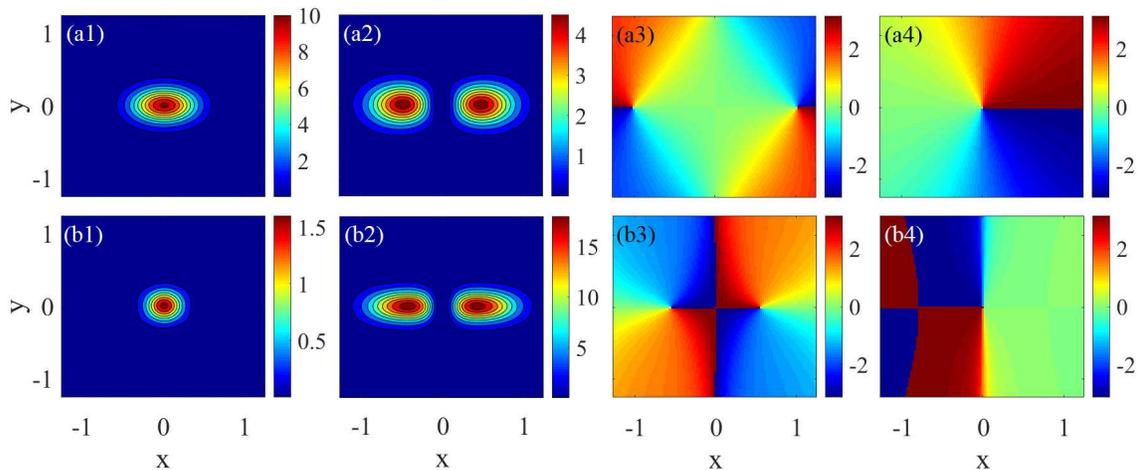}}
\caption{The same as in Fig. \protect\ref{SVexp}, but for the inverted AVSs,
with $\Omega <0$. The first row represents a stable AVS with $(N,\Omega
)=(6,-1.82)$, with equal norms in the fundamental and vortex components. The
second row shows the soliton with $(N,\Omega )=(10,-10.8)$, which is a
dipole mode for $\protect\phi _{-}$ and located at the stability boundary
[see Fig. \protect\ref{FM}(a) below].}
\label{SVexp2}
\end{figure*}

\begin{figure*}[t]
{\includegraphics[width=1.2\columnwidth]{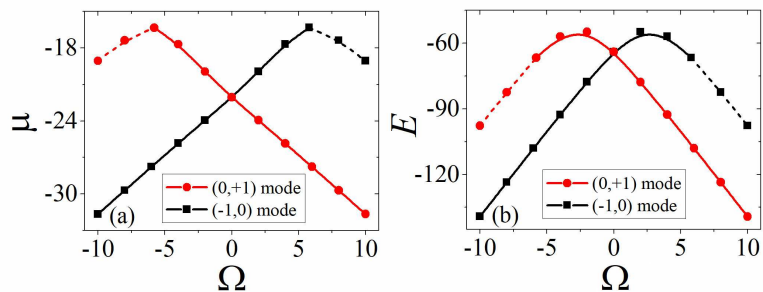}}
\caption{The chemical potentials (a) and energy (b) for the mode under the
consideration, with vorticities in the $\protect\psi _{+}$ and $\protect\psi %
_{-}$ components $\left( S_{+},S_{-}\right) =(0,+1)$, and its flipped
counterpart, with $\left( S_{+},S_{-}\right) =\left( -1,0\right) $, with a
fixed norm, $N=8$, vs. the Zeeman-splitting strength, $\Omega $.}
\label{muESV2}
\end{figure*}

\begin{figure*}[t]
{\includegraphics[width=1.8\columnwidth]{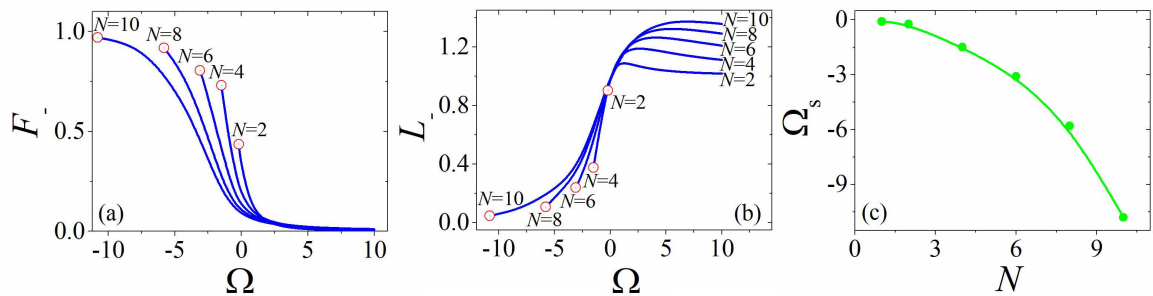}}
\caption{(a) The share of the total norm in the vortex component [see Eq. (%
\protect\ref{F})], $F_{-}(\Omega )$, as a function of the Zeeman-splitting
strength, $\Omega $, at fixed values of the total norm, $N$. (b) The angular
momentum of the vortex component, defined as per Eq. (\protect\ref{Leq}),
vs. $\Omega $, at fixed values of $N$. In both panels (a) and (b), stable
branches are shown by blue solid curves. Red circles at the end of the blue
curves designate the stability boundary, $\Omega =\Omega _{\mathrm{s}}(N)$.
The inverted modes (corresponding to $\Omega <0$) exist too at $\Omega
<\Omega _{\mathrm{s}}(N)$, but are unstable in that region. (c) The
stability boundary, $\Omega _{\mathrm{s}}$, as a function of $N$. }
\label{FM}
\end{figure*}

\begin{figure*}[t]
{\includegraphics[width=1.2\columnwidth]{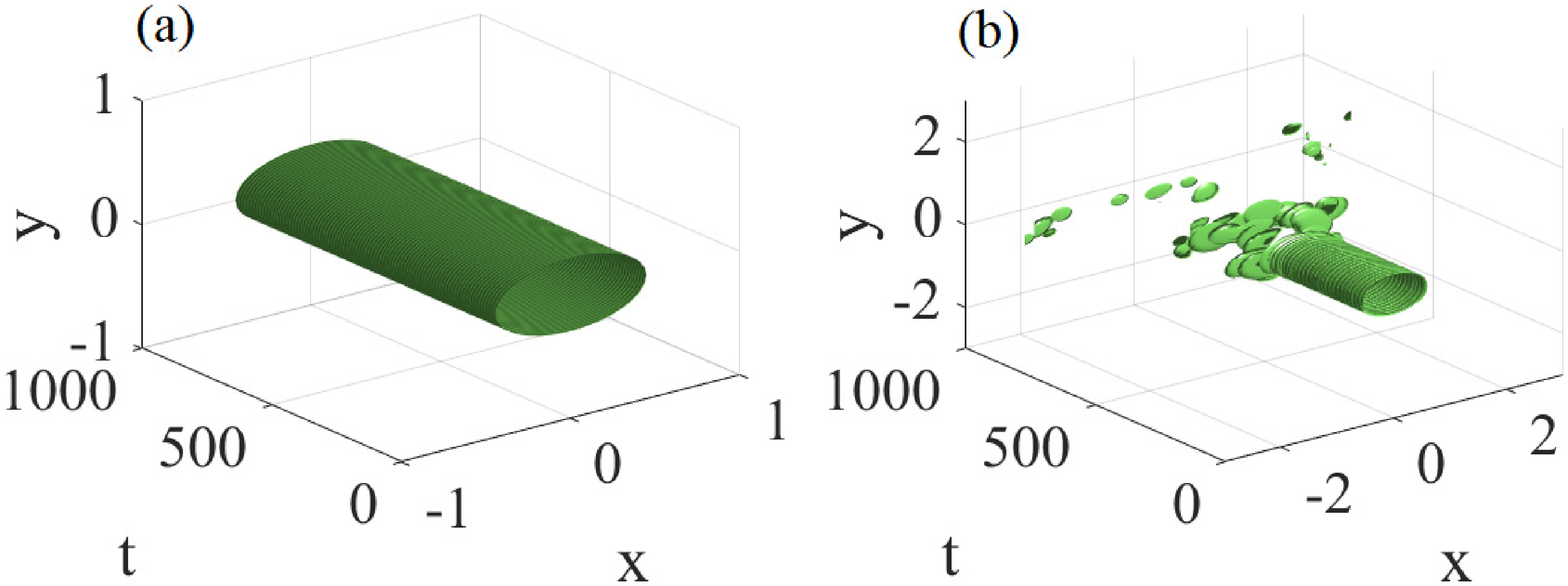}}
\caption{Plots of $|\protect\phi _{+}(x,y,t)|^{2}$ depict evolution of
typical examples of stable (a) and unstable (b) inverted AVSs, for $%
(N,\Omega )=(5,-1.5)$ and $(5,-2.4)$, respectively. For this value of the
total norm, $N=5$, the stability boundary is $\Omega _{\mathrm{s}}=-2.2$,
see Fig. \protect\ref{FM}(c). }
\label{Expstableunstable}
\end{figure*}

As stressed above, we considered the soliton modes of semi-vortex type, with
their fundamental and vortex constituents sitting, respectively, in lower-
and higher-energy components, in terms of the ZS with $\Omega >0$. Here, we
aim to consider the reverse setting, with $\Omega <0$, when the fundamental
and vortex parts find themselves in the higher- and lower-energy components,
respectively. Because the fundamental constituents tends to carry a much
larger share of the total norm than its vortical counterpart, one may expect
that the inverted state will be an excited one, while the stable states
considered above were ground states in the system. For this reason, it is
quite interesting to explore the shape and, especially, stability of these
excited states. Surprisingly, it is found below that they are stable in a
considerable part of their parameter space.

Generic examples of the density and phase structure of the inverted AVSs are
displayed in Fig. \ref{SVexp2}. As well as these examples, all the inverted
AVSs feature the horizontal structure, as it is defined above (with two
density maxima separated along the $x$ axis, i.e., along the direction of
the polarization of dipoles in the underlying condensate). Unlike the case
of large positive $\Omega $ considered above, vertical inverted AVSs have
not been found.

The dependence of the AVS's chemical potential and energy on $\Omega $ at
fixed norm $N$ is displayed in Fig. \ref{muESV2}, both for the mode under
the consideration, and its \textit{flipped} counterpart, with vorticities $%
-1 $ and $0$ in components $\psi _{+}$ and $\psi _{-}$, respectively. In the
absence of the ZS, $\Omega =0$, both semi-vortex modes are fully equivalent
to each other \cite{SVS1}. Further, the present mode at given $\Omega $ and
the flipped one at the ZS strength $-\Omega $ remain equivalent too. Thus,
Fig. \ref{muESV2}(b) makes it obvious that the present mode is an excited
state at $\Omega <0$, as its energy is higher than the energy of the ground
state, which is realized, at $\Omega <0,$ by the flipped AVS\ mode.

Further, Fig. \ref{FM}(a) represents families of the AVS modes by displaying
the share of the norm in the vortex component, $F_{-}$, defined according to
Eq. (\ref{F}), vs. $\Omega $ at several fixed values of $N$. Naturally, $%
F_{-}$ increases with the growth of $-\Omega $. A nontrivial finding is
that, although the inverted AVS is an excited state, as shown above, it has
a vast stability region, $\Omega >\Omega _{\mathrm{s}}(N)$, with the
dependence of the stability boundary on $N$, i.e., $%
\Omega _{\mathrm{s}}(N)$, shown in Fig. \ref{FM}(c).  In particular, Fig. \ref{FM}(a) demonstrates that,
at $N>2$, the AVS stabilizes itself by shifting more than half of the total
norm to the lower-energy vortex component, making it a dominant one in the
semi-vortex soliton. In previously studied models, stable SVs with a
dominant vortex component were never found \cite%
{SVS1,SVS3,QD2017,Xunda2016,SOCgapsoliton}. In fact, previous studies have
never produced \emph{stable} excited states of SVs either (unstable excited
states were found in some cases \cite{SVS1,QD2017}).

As the vortex mode becomes dominant with the increase of $-\Omega $ (still
remaining stable), its shape actually becomes \textquotedblleft less
vortical", as may be quantified by the dependence of its angular momentum,
which is calculated pursuant to Eq. (\ref{Leq}) and displayed, as a function
of $\Omega $, in Fig. \ref{FM}(b). The plots displayed in the second row of
Fig. \ref{SVexp2} suggest that the trend to the loss of the angular momentum
with the growth of $-\Omega $ (at fixed $N$) may be explained by gradual
evolution of the dominant component toward a dipole mode [See in the second
row of Fig. \ref{SVexp2}], while keeping the vorticity.

Typical examples of a stable and unstable inverted AVSs, which are selected,
respectively, from intervals $\Omega >\Omega _{\mathrm{s}}$ and $\Omega
<\Omega _{\mathrm{s}}$ for the same $N$, are displayed in Fig. \ref%
{Expstableunstable}. The unstable mode suffers spontaneous fragmentation
after stably propagating for a while.

\section{Conclusion}

The objective of this work is to study anisotropic vortex solitons (AVSs) in
dipolar SO (spin-orbit)-coupling BECs, controlled by the strength of the ZS
(Zeeman splitting). The dipolar condensate was considered in the strongly
anisotropic form, with the in-plane polarization of the atomic dipoles. This
setting, although being relatively complex, makes it possible to construct
stable anisotropic vortices [in fact, \textit{semi-vortices} (SVs), i.e.,
bound states of zero-vorticity and vorticity-$1$ components], which was not
possible in previously studied systems.

Here, we have demonstrated that the shape and stability of the AVSs are
efficiently controlled by the ZS strength, $\Omega $. In the case of $\Omega
>0$, which corresponds to the AVS's fundamental and vortex constituents
placed, respectively, in the lower- and higher-energy components, as defined
by the ZS, the increase of $\Omega $ leads to the transition from the
horizontal to vertical shape of the AVS, the horizontal direction being the
one aligned with the atomic dipole moments. The value of $\Omega $ at the
horizontal-vertical transition, $\Omega _{\mathrm{C}}$, was found in the
numerical form, and its nearly linear dependence on the total norm, $N$, was
explained. Exactly at $\Omega =\Omega _{C}$, the AVS assumes an
elliptic-ring shape.

Mobility of the AVSs was studied too, by means of direct simulations of
initially kicked solitons. As a result, they exhibit circular motion with a
small radius, with a systematic drift in the direction of the kick or \emph{%
against} it. It was found that the AVS's effective mass is \emph{negative}
in an interval of $0<\Omega <\Omega _{M=\infty }$, and positive at $\Omega
>\Omega _{M=\infty }$. The effective mass changes its sign at $\Omega
=\Omega _{M=\infty }$ via the divergence, as $M_{\mathrm{eff}}\sim \left(
\Omega -\Omega _{M=\infty }\right) ^{-1}$.

Unexpected results have been reported for the \textit{inverted} AVSs, i.e.,
ones at $\Omega <0$, with the fundamental and vortex constituents placed by
the ZS in the higher- and lower-energy components, respectively. The
inverted AVS is always an excited state (with a horizontal structure), which
coexists with a lower-energy ground state (horizontal or vertical one). A
surprising result is that these excited states are stable in a large
parameter region. In previously studied SO-coupled nonlinear systems,
excited were always unstable.

The present analysis can be extended in some directions. First, a natural
possibility is to simulate collisions between mobile AVSs. Next, one can
consider the limit case of strong SO coupling and strong ZS, when the
kinetic-energy terms may be neglected in Eq. (\ref{fulleq}), which gives
rise to the bandgap spectrum \cite{SOCgapsoliton}. It may be interesting to
construct AVSs of the gap-soliton type, that may populated the bandgap. On
the other hand, it may be interesting too to consider the ZS with a
time-dependent strength and the respective \textquotedblleft management"
regimes for AVSs. Finally, a challenging option is to seek extension of
current setting to the 3D geometry.


\begin{acknowledgments}
This work was supported, in part, by NNSFC (China) through Grant No.
11575063, 61471123, 61575041, by the joint program in physics between NSF and Binational
(US-Israel) Science Foundation through project No. 2015616, and by the Natural Science Foundation of Guangdong
Province, through Grant No. 2015A030313639. B.A.M. appreciates a foreign-expert grant of the Guangdong province (China).
\end{acknowledgments}

\end{document}